\documentclass[aps,prc,amsmath,amssymb,twocolumn,showpacs,floatfix]{revtex4}
\usepackage{epsfig,graphics,color}
\usepackage{graphicx}% Include figure files
\usepackage{dcolumn}% Align table columns on decimal point
\usepackage{bm}% bold math
\usepackage{amsmath}
\newcommand \etc {{\it etc.}}
\newcommand \tie {{\it i.e.}}

\newcommand \kd  {\delta}
\newcommand \ra  {\rightarrow}

\newcommand \w  {\omega}

\newcommand \g {\gamma}
\newcommand \ro {\rho}

\newcommand \e {\epsilon}

\newcommand \B {\beta}

\newcommand \lc {\langle}
\newcommand \rc {\rangle}
\newcommand \prt {\partial}
\newcommand \D {\Delta}
\newcommand \sg {\sigma}

\newcommand \nt {\noindent}

\newcommand \bvec{\left( \begin{array}{c} }
\newcommand \evec{\end{array} \right)}

\newcommand \eg {{\it e.g.}}
\newcommand \bea{\begin{eqnarray} }
\newcommand \eea{\end{eqnarray} }
\newcommand \nn {\nonumber}
\newcommand {\be} {\begin{equation}}
\newcommand {\ee} {\end{equation}}

\newcommand{\sq}{\sigma^2}
\newcommand{\hn}{\hat{N}}
\newcommand {\ub}{\bar{u}}
\newcommand {\db} {\bar{d}}

\begin{document}

\title{Baryonic Strangeness and Related Susceptibilities in QCD}

\author{A. Majumder}
\affiliation{Department of Physics, Duke University, Durham, NC 27708-0305, USA}

\author{B. M\"{u}ller}
\affiliation{Department of Physics, Duke University, Durham, NC 27708-0305, USA}

\date{ \today}

\begin{abstract}
The ratios of off-diagonal to diagonal conserved charge susceptibilities 
\eg, $\chi_{BS}/\chi_{S}, \chi_{QS}/\chi_{S}$, related  to the quark flavor 
susceptibilities, have proven to be discerning probes of the flavor carrying 
degrees of freedom in hot strongly interacting matter. Various constraining 
relations between the different  susceptibilities are derived based on the 
Gell-Mann-Nishijima formula and the assumption of isospin symmetry. Using generic 
models of deconfined matter and results form lattice QCD, it is demonstrated 
that the flavor carrying degrees of freedom at a temperature above $1.5T_c$ 
are quark-like quasiparticles. A new observable related by isospin symmetry 
to $C_{BS}=-3\chi_{BS}/\chi_{S}$ and equal to it in the baryon free regime
is identified. This new observable, which is blind to neutral and non-strange 
particles, carries the potential of being measured in relativistic heavy-ion 
collisions.
\end{abstract}

\pacs{12.38.Mh, 11.10.Wx, 25.75.Dw}

\maketitle

%%%%%%%%%%%%%%%%%%%%%%%%%%%%%%%%%%%%
%%%%%%%%%%%%%%%%%%%%%%%%%%%%%%%%%%%%
%%%%%%%%%%%%%%%%%%%%%%%%%%%%%%%%%%%%
%%%%%%%%%%%%%%%%%%%%%%%%%%%%%%%%%%%%

\section{Introduction}
%%%%%%%%%%%%%%%%%%%%%%%%%%%%%%%%%%%%
%%%%%%%%%%%%%%%%%%%%%%%%%%%%%%%%%%%%
%%%%%%%%%%%%%%%%%%%%%%%%%%%%%%%%%%%%
%%%%%%%%%%%%%%%%%%%%%%%%%%%%%%%%%%%%

The goal of of the heavy-ion program at the Relativistic Heavy Ion Collider 
(RHIC)  is the creation and study of heated strongly interacting matter at 
nearly vanishing baryon density \cite{Harris:1996zx,Gyulassy:2004zy}. 
Detailed models of nuclear reactions predicted that the energy deposition
in the center-of-mass frame should be sufficient to cause temperatures 
at mid-rapidity in central collisions of gold nuclei to reach upwards 
of 300 MeV\cite{Wang:1996yf}. These predictions have been confirmed by 
the experimental results of the four RHIC detector collaborations, which 
set a lower bound of about 5 GeV/fm$^3$ on the energy density at a time 
$\tau = 1$ fm/c in central Au+Au collisions \cite{RHIC_Whitepapers}.
This estimate for the attained energy density should place the produced 
matter well into region of the QCD phase diagram that cannot be described
as a dilute hadronic resonance gas. Until recent results from the RHIC 
experiments have raised doubt about this interpretation \cite{Gyulassy:2004zy},
the matter in this domain of the QCD phase diagram was expected to be a
colored plasma composed of quasiparticle excitations with the quantum 
numbers of quarks and gluons \cite{Harris:1996zx,Rischke:2003mt}.

Signals of the excited matter produced in the early stages of such a collision, 
buried in the pattern of detected particles, may be divided into two categories: 
hard probes, such as the modification of partonic jets by the medium 
\cite{Gyulassy:2003mc}, bound states of heavy quarks \cite{Satz:1995sa}, 
and electromagnetically produced particles \cite{Gale:2003iz}; and bulk 
observables, dealing with low momentum particles which make up a large 
fraction of the produced matter. Bulk observables include the single 
inclusive spectra of identified hadrons \cite{Braun-Munzinger:2003zd} and 
the event-by-event fluctuations of conserved charges \cite{Asakawa:2000wh}. 
The latter are the subject of our present study.  

The theoretical analysis of the observed suppression of energetic hadron
emission in Au+Au collisions at RHIC (``jet quenching'' \cite{Gyulassy:2003mc}) 
confirms that matter with a very high energy density is produced. This matter 
clearly exhibits collective behavior as evidenced by its radial and elliptic 
flow \cite{Ackermann:2000tr}. As a large elliptic flow requires large pressure 
gradients which can only be present during the earliest stage of the collision, 
the matter produced at such times must exhibit the properties of a fluid. 
A quantitative analysis of the observed magnitude of the flow indicates that
this fluid must be nearly ideal, i.~e.~endowed with a very low viscosity
\cite{Teaney:2003kp}. This result suggests that there must exist a strong 
interaction between the constituents of the medium. 

At vanishing baryon density, the entire range of thermal conditions expected 
to be attained at RHIC may be simulated by the numerical methods of lattice 
QCD (LQCD) at finite temperature  \cite{Karsch:2003jg}. Here one computes the 
grand canonical partition function of a system whose states are thermally 
weighted by the QCD action at a temperature $T$, with baryon, electric charge, 
and strangeness chemical potentials set to zero ($\mu_B = \mu_Q = \mu_S = 0$).
Explorations on the lattice consist of a three-fold approach \cite{Karsch:2003jg}: 
studies of the behavior of the components of the stress energy tensor, i.~e.~the 
energy density $\e$ and the pressure $P$; spatial and temporal correlation 
functions, and the recently measured derivatives of the free energy such as 
the various conserved flavor susceptibilities. 

Investigations of the first kind, 
exploring the QCD equation of state at vanishing chemical potentials, provide
the most solid evidence for the expectation that, when hadronic matter is 
heated beyond a critical temperature $T_c \sim 170$MeV, a transition to 
a new state of matter, the quark-gluon plasma (QGP), will occur. The 
transition is signalled by a steep rise in the energy density and the pressure 
as a function of the temperature. The slow rise of both quantities prior to the 
sudden transition has come to be understood in a picture of a hadronic 
resonance gas \cite{Karsch:2003zq}. However, attempts at an description of 
the excited phase as a weakly interacting plasma of quasiparticles 
\cite{Andersen:1999va,Blaizot:2003iq} has not met with success in the region 
from $T_c \leq T \leq 3T_c$. 

This finding might indicate that matter in this region may not be a weakly 
coupled plasma where quarks and gluons are deconfined over large distances 
as it was originally proposed \cite{Collins:1974ky}. It has been established 
that such a weakly coupled state, indeed, occurs at much higher temperatures 
(see \cite{Blaizot:2003tw} and references therein). Assuming that temperatures 
at RHIC do not exceed $3T_c$, a strongly coupled state is not inconsistent 
with the strong collective behavior observed in experiments. It is clear that 
a microscopic understanding of the emergent degrees of freedom in this regime 
is essential for an explanation of the ``perfect liquid'' character of the 
matter created in nuclear collisions at RHIC. 

There already exists a candidate model for such matter, proposed by Shuryak 
and Zahed \cite{Shuryak:2004cy,Shuryak:2004tx}, consisting of a tower of 
colored bound states of heavy quasi-particulate quarks and gluons. In spite 
of the presence of such heavy quasi-particles, this model was to account for 
the large pressure required by the RHIC data because of the proliferation of 
excited bound states. The large cross sections resulting from resonance 
scattering were put forward as the cause for the short mean free path 
which is a requisite for the appearance of hydrodynamic behavior. The model 
has, however, fared poorly in a comparison with lattice susceptibilities. 
Koch {\em et al.}~\cite{Koch:2005vg} recently proposed the susceptibilities 
of conserved charges $B$ (baryon number), $Q$ (electric charge) and $S$ 
(strangeness) and their off diagonal analogues as diagnostics of the conserved 
charge or flavor carrying degrees of freedom in a strongly interacting system.
The primary quantity of interest is the ratio of the covariance between 
baryon number and strangeness $\sg_{BS}^2  \propto \chi_{BS}$ to the variance 
in strangeness $\sg_{S}^2 \propto \chi_S $, renormalized to be unity in a 
quasi-particle gas of quarks and gluons, 
\bea
C_{BS} =  -3 \frac{\lc \kd B \kd S \rc}{\lc \kd S^2 \rc} 
       = -3 \frac{\chi_{BS}}{\chi_{S}}.
\eea
This quantity may be calculated on the lattice and estimated in the dynamical 
model of ref.~\cite{Shuryak:2004cy}. The calculated values on the lattice were 
found to be 50\% higher than those computed in the model \cite{Koch:2005vg}. 
Such comparisons demonstrate the capacity of  ratios such as $C_{BS}$ to 
serve as tests for models of the QGP.  The origin of the difference between
the results from lattice QCD and those from the bound state model will be 
further discussed in the upcoming sections. 

The objective of the present work is to continue the study of diagonal and 
off-diagonal flavor susceptibilities. In the next section, we point out that 
there exists an entire gamut of such diagonal and off-diagonal susceptibilities
for conserved quantum numbers, and they are related by a simple transformation 
to the equivalent basis of quark flavor susceptibilities. This presentation 
builds on the work of Gavai and Gupta~\cite{Gavai:2005yk}. In the next 
section, a complete set of such susceptibilities is introduced and related 
via the Gell-Mann-Nishijima formula.  The alternative set of quark flavor 
susceptibilities, which appear more often in the literature, is also compared. 
In Section III operator relations between the off-diagonal susceptibilities 
in different bases are derived and the generic behavior of such operators 
depending on the prevalent degrees of freedom is outlined.  In Section IV 
several models for the flavor carrying sector of the QGP are analyzed, and 
arguments favoring a picture of quasi-particle quarks are presented. In 
Section V we formulate observables based on the ratios of susceptibilities 
that may be estimated from experimental measurements. Concluding discussions 
are found in Section VI.

%%%%%%%%%%%%%%%%%%%%%%%%%%%%%%%%%%%%

\section{Diagonal and off-diagonal susceptibilities}

%%%%%%%%%%%%%%%%%%%%%%%%%%%%%%%%%%%%

In the lattice formulation of QCD, the fundamental degrees of freedom
are local quark and gluon fields. Under conditions where deconfinement 
has been achieved, the elementary set of conserved charges is given by
the quark flavors: the net ``upness'' ($\D u = u - \bar{u}$), 
``downness'' ($\D d = d - \bar{d}$) and ``strange-quarkness'' 
($\D s = s - \bar{s} $). An alternate basis is provided by the hadronically 
defined conserved charges of baryon number ($B$), electric charge ($Q$) 
and strangeness ($S$). The two bases are related by
\bea
B &=& \frac{1}{3} (\D u + \D d + \D s), \nn \\
Q &=& \frac{2}{3}\D u - \frac{1}{3} \D d - \frac{1}{3} \D s ,\nn \\
S &=& - \D s. \label{mean_mat}
\eea
In what follows, the ``$\D$'' will be omitted and the variables $u,d,s$ 
will be understood to denote the net flavor contents. 

The mean values of any conserved charge may be measured in a thermal 
ensemble of interacting quarks and gluons on the lattice. The grand 
canonical partition function may be defined using either basis \tie,
\bea
\mathcal{Z}(T,\mu_B,\mu_Q,\mu_S) = \mathcal{Z}(T,\mu_u,\mu_u,\mu_s) ,\nn \\
\mbox{if} \,\,\, \mu_u  = \frac{\mu_B}{3} + \frac{2\mu_Q}{3} \, \& \,\,
\mu_d = \frac{\mu_B - \mu_Q}{3}
\,\&\,\, \mu_s = -\mu_S.
\eea
The mean values and variances of any combination of conserved charges may be
obtained from appropriate differentiation of either partition function:
\bea
\lc x \rc = T \frac{ \prt }{\prt \mu_x} \mbox{log}\mathcal{Z}(T,\mu_x,\mu_y) , 
\label{mean} \\
 \sq_{xy} = T^2 \frac{\prt^2}{\prt \mu_x \prt \mu_y} \log{\mathcal{Z}(T,\mu_x,\mu_y)}. \label{sig2}
\eea
While the mean values in a given basis are related to the other via 
Eq.~\eqref{mean_mat}, the variances exhibit a more complex structure. 
A similar and larger matrix relates the six fluctuation measures, viz.\ 
the variances $\sq_B,\sq_Q,\sq_S$ and the covariances $\sq_{BS},\sq_{BQ},\sq_{QS}$, 
to the six diagonal and off-diagonal quantities constructed from the quark 
flavors. The  $6 \times 6$ matrix relating these two sets of (co-)variances
is given by 
\bea 
\left(
\begin{array}{c}
B^2\\Q^2\\BQ\\BS\\QS\\S^2
\end{array}
\right)
= \frac{1}{9}
\left[
\begin{array}{rrrrrr}
 1 &  1 &  2 &  2 &  2 &  1 \\
 4 &  1 & -4 & -4 &  2 &  1 \\
 2 & -1 &  1 &  1 & -2 & -1 \\
 0 &  0 &  0 & -3 & -3 & -3 \\
 0 &  0 &  0 & -6 &  3 &  3 \\
 0 &  0 &  0 &  0 &  0 &  9
\end{array}
 \right]
\left(
\begin{array}{c}
u^2\\d^2\\ud\\us\\ds\\s^2
\end{array}
\right),
\eea
\nt
where the corresponding subscripts of $\sq_{xy}$ are used
to indicate the corresponding variance. The above matrix 
immediately demonstrates the utility of using ratios of 
$\sq_{BS},\sq_{QS}$ and $\sq_{S}$ as opposed to the 
other three variances, as these form a smaller subgroup with 
the quark flavor covariances $\sq_{us},\sq_{ds}$ and the 
strangeness variance $\sq_{s}$.
The (co-)variances are extensive quantities: 
\be
\sq_{xy} = VT \chi_{xy}, 
\ee 
where $\chi_{xy}$ is the intensive diagonal or off-diagonal susceptibility.
These susceptibilities can be measured on the lattice. In heavy-ion experiments, 
the variances and covariances are measured by means of an event-by-event 
analysis of the corresponding conserved quantities, \tie,
\bea
\sq_{xy} = \frac{1}{N_{E}}\sum_{i \in E} X_i Y_i - 
\left(\frac{\sum_{i \in E} X_i}{N_{E}}\right) 
\left(\frac{\sum_{i \in E} Y_i}{N_{E}}\right),
\eea
where $E$ represents the set of events, $N_E$ is the number of events
considered, and $X_i,Y_i$ are the net values of the conserved charge in 
a given event $i$. The volume independent ratios of variances, measured 
event-by-event in heavy-ion collisions, may then be directly compared 
with the lattice estimate for the ratio of susceptibilities. 

Besides relating the susceptibilities or variances from one basis to 
another, simplifying relations may be obtained between the variances in 
a given basis using the Gell-Mann-Nishijima formula, 
\begin{equation}
Q = I_3 + \frac{B + S}{2},
\end{equation}
where, $I_3$ denotes the third component of the isospin. Because the mass 
difference between the $u,d$ quarks is small compared with the typical
scale of hadron masses, the masses of all hadrons in a given isospin 
multiplet are degenerate. One may also assume that any quasi-particle
excitations in highly excited chromodynamic matter display a large degree 
of isospin symmetry. It should be pointed out that in actual experiments, 
besides a small mass split between the $u$ and the $d$ quark, there is an 
overall isospin asymmetry in the entire system as all heavy nuclei have 
a large isospin due to their proton-neutron imbalance. Such an isospin 
asymmetry is, however, dominantly correlated with the net baryon number 
of the nuclei. As the matter formed in the central rapidity region has 
nearly vanishing net baryon density, one may assume that the produced 
matter has almost no net $I_3$ component, on the average. In the following, 
we will also ignore the isospin violating effects of weak decays and 
electromagnetic interactions. 

To obtain relations between quadratic variables involving strangeness, 
we obtain the variance of the left and right hand side of the equation
\begin{eqnarray}
QS &=&  I_3 S + \frac{BS + S^2}{2}.
\end{eqnarray}
Taking the ensemble or event average of the above quantity, we obtain
\begin{eqnarray}
\sq_{QS} &=& \frac{1}{N_E} \left[ \sum_{i \in E} Q_i S_i  
- \frac{1}{N_E} \sum_{i \in E} Q_i \sum_{j \in E} S_j  \right] \nn \\
&=& \left[ \frac{1}{N_E} \sum_{i \in E} \sum_f n_i^f Q_f  \sum_g n_i^g S_g \right. 
\nn \\
 &-&\left.  \frac{1}{N_E} \sum_{i \in E} \sum_f n_i^f Q_f  
%\\
\frac{1}{N_E}\sum_{j \in E} \sum_g n_j^g S_g \right] .  \label{sg_qs}
\end{eqnarray}
In the above,  the total charge and strangeness measured in an event $i$ is 
denoted as $Q_i,S_i$. In the case that the degrees of freedom or active 
flavors $f$  are eigenstates of charge or strangeness, this expression may 
be decomposed as $Q_i = \sum_f n_i^f Q_f $ where $n_i^f$ are the number of 
states of flavor $f$ in event  $i$. For independent flavors, where 
$\lc n^f n^g \rc = \sum_i^{Ens.} n_i^f  n_i^g =  \lc n^f \rc \lc n^g \rc$, one 
may easily demonstrate that
\bea
\sq_{QS} &=& \sum_f \sq_f  Q_f  S_f  =\!\!\!\!\!\!\!\!\mbox{}^{P.S.} \sum_f  \lc n^f \rc Q_f S_f. \label{sg_qs2}
\eea
The last equality in the above equation holds solely in the case that Poisson 
statistics is applicable to the independent flavors (\tie, the variance $\sq$ 
is equal to the mean $\lc n \rc$) \eg, in the case  where the masses of the 
various flavors exceeds the temperature.  

If isospin symmetry is maintained by the system then $\sq_f = \sq_g$ when $f,g$ belong
to the same isospin multiplet. The associated physical picture is that fluctuations of isospin, 
or of a product quantum number involving isospin are brought about by fluctuations in the 
populations of flavors that carry isospin. If all the members of an isospin multiplet have 
the same mass and there exist no chemical potentials which favor one species over another, 
the fluctuations of carriers with opposing values of $I_3$ compensate for each other. 
As a result , one obtains the equalities, 
\begin{equation}
\sum_{i=-I}^{I} {I_3}_i \sq_i \simeq  \sum_{i=-I}^{I} {I_3}_i S_i \sq_i \simeq \sum_{i=-I}^{I} {I_3}_i B_i \sq_i \simeq 0,
\label{isospin}
\end{equation}
where, the sum is restricted to lie in a given isospin multiplet.
One uses the notation of an approximate equality ($\simeq$) as 
opposed to an exact equality ($=$) to high-light the fact that such 
relations hold only in the case of exact isospin invariance.
%  $m_u = m_d$.
However, even in the case with a small mass split between the  $u$ and 
$d$ quark, the equalities are still approximately true  and will 
be applied in this spirit. In actual calculations involving quarks and 
hadrons, in the upcoming sections,  the physical masses of all known 
species will be used.
Based on the above, the following simplifying relations 
between the various covariances may be easily derived,
\bea 
\sq_{QS} &=& \sq_{I_3 S } + \frac{ \sq_{BS} + \sq_S} {2} \simeq \frac{\sq_{BS} + \sq_S}{2} \label{s_qs_and_bs} \\
\sq_{QB} &=& \sq_{I_3 B } + \frac{ \sq_{B} + \sq_{BS}} {2} \simeq \frac{ \sq_{B} + \sq_{BS}  } {2} 
\label{s_qb_and_sb}
\eea
In the definitions introduced in Refs.~\cite{Koch:2005vg,Gavai:2005yk}, 
for the two coefficients: 
$C_{BS} = -3 \sq_{BS}/\sq_S$  and $C_{QS} = 3 \sq_{QS}/\sq_S$, one may 
use Eq.~\eqref{s_qs_and_bs} to obtain the following simplifying relation,
\begin{equation}
C_{QS} \simeq \frac{3 - C_{BS}}{2} \label{cqs_cbs}
\end{equation}

\begin{figure}
\resizebox{2.7in}{2.7in}{\includegraphics[0in,0in][5in,5in]{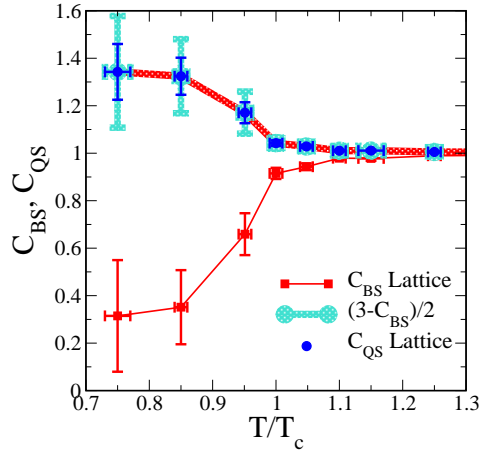}}
\caption{(Color Online) A test of the formula $C_{QS} = \frac{3 - C_{BS}}{2}$. 
Both ratios, $C_{BS}$ and $C_{QS}$, computed on the lattice in 
Ref.~\cite{Gavai:2005yk}, are presented as the solid red and blue points. 
The $C_{BS}$ points are then used to calculate $C_{QS}$ resulting in the 
hazed cyan points.}
\label{fig1}
\end{figure}

The validity of the above equation is amply demonstrated by Fig.~\ref{fig1}. 
The solid red and blue data points are taken form the lattice calculations 
of Ref.~\cite{Gavai:2005yk}, where both these quantities were computed 
independently of each other.  The hazed cyan points represent an estimation 
of $C_{QS}$ from the $C_{BS}$ points using Eq.~\eqref{cqs_cbs}. In the 
lattice computation, the masses of the $u$ and $d$ quarks are set equal 
to each other, \tie, the lattice calculation displays exact isospin symmetry, 
clearly demonstrated by the exact coincidence of the hazed-cyan and solid-blue circles. 
It should be reiterated that $C_{QS}$ may be calculated given a $C_{BS}$ 
using Eq.~\eqref{cqs_cbs}.  

Similarly, one may also define another set of related observables, 
\bea
C_{QB} &=& \frac{\sq_{QB}}{\sq_{B}} , 
C_{SB} = \frac{\sq_{SB}}{\sq_{B}}.
\eea
We refrain from ascribing any overall normalization factor, as in the 
cases of $C_{BS},C_{QS}$:  none is self-evident. Both these 
quantities may be estimated on the lattice, in models, as well as 
in an experiment. In all such cases, independent of the phase of 
matter involved, one will recover the equality imposed by isospin 
symmetry (Eq.~\eqref{s_qb_and_sb}), 
\bea
C_{QB} = \frac{1+C_{SB}}{2}. \label{cqb_csb}
\eea
An interesting situation is afforded for flavor SU(2), \tie, in a theory 
without strangeness. While $C_{BS}, C_{SB}$ and $C_{QS}$  are undefined, 
$C_{QB} = \frac{1}{2}$ (a similar situation is that of the canonical ensemble 
where $S$ is held fixed) .  This last value serves as an important test of 
any model devised to reproduce the properties of any phase of strongly 
interacting matter and will be used in the upcoming  sections where we will 
compare model predictions to results from lattice simulations. 

%%%%%%%%%%%%%%%%%%%%%%%%%%%%%%%%%%%%
%%%%%%%%%%%%%%%%%%%%%%%%%%%%%%%%%%%%
%%%%%%%%%%%%%%%%%%%%%%%%%%%%%%%%%%%%
%%%%%%%%%%%%%%%%%%%%%%%%%%%%%%%%%%%%
\section{Operator Relations}
%%%%%%%%%%%%%%%%%%%%%%%%%%%%%%%%%%%%
%%%%%%%%%%%%%%%%%%%%%%%%%%%%%%%%%%%%
%%%%%%%%%%%%%%%%%%%%%%%%%%%%%%%%%%%%
%%%%%%%%%%%%%%%%%%%%%%%%%%%%%%%%%%%%

In the previous section, we outlined various properties and relations 
between the diagonal and off-diagonal susceptibilities of heated strongly 
interacting matter.  In this section, we extend the discussion to the 
operator structure of these susceptibilities and derive general expectations
for the value of the off-diagonal flavor susceptibilities in different 
quasi-particle bases. The reader not interested in such a study can skip 
to the next section, in which comparisons of lattice results with 
phenomenological models will be made. 

\begin{figure}
\resizebox{2.7in}{2.7in}{\includegraphics[0.0in,0.0in][5.5in,5.5in]{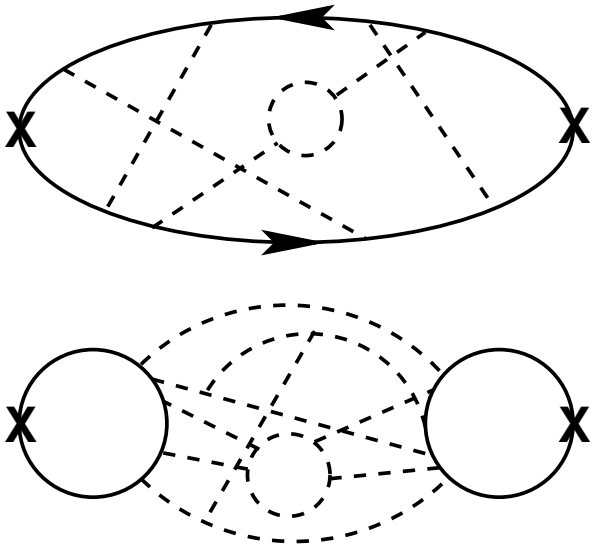}}
\caption{Connected and disconnected quark number operators.}
\label{fig2}
\end{figure}

In numerical simulations of lattice QCD, one evaluates thermal expectation 
values of operators weighted with the SU(3) gauge action $S_g$ and the 
fermionic determinant det$[M]$,  
\bea
\lc \mathcal{O} \rc = \frac{\int \mathcal{D} U   \mathcal{O} 
(\mbox{det}[M])^{n_f} e^{-S_g} }
{\int \mathcal{D} U  (\mbox{det}[M])^{n_f} e^{-S_g}}
\eea
In the case of the quark number susceptibility, the operator in question is
\bea
N_{q_i} N_{q_j} = \int d^3 x  d^3 y n_{q_i}(x) n_{q_j}(y) ,
\label{sus_oprtr}
\eea
where $n_{q_i}(x) = : \bar{\Psi}_i(x) \g^0 \Psi_i (x) :$  and $q_{i}, q_{j}$ 
represent quarks of flavor $i$ and $j$. The normal ordering removes the 
leading short distance piece, which is proportional to the four-volume. 
The operator may be decomposed into connected and disconnected diagrams 
as shown in Fig.~\ref{fig2}. The locations $x,y$ of the two operator 
insertions are indicated by the crosses in the figures. The solid lines 
represent the valence quarks, whereas the dashed lines represent virtual 
gluons or quarks as allowed by the Lagrangian. If the flavors of the two 
quarks are the same ($i=j$), corresponding to a diagonal susceptibility, 
contributions emerge from both types of diagrams. Whereas if the two 
flavors are different, contributions arise solely from the second diagram. 

In the interaction picture, the operator of Eq.~\eqref{sus_oprtr}, for the 
off-diagonal susceptibility may be expressed in the simplified form,
\bea
N_{q_i} N_{q_j} &=& \int d^3 q_i \int d^3 q_j   
\sum_{r,s} \left[  {a^r_i}^\dag a^r_i - {b^r_i}^\dag b^r_i  \right]  \nn \\
 &\times&  \left[  {a^s_j}^\dag a^s_j - {b^s_j}^\dag b^s_j  \right],
\eea
where $r,s$ denote the spin orientations.  This expression may then be 
evaluated in a basis of weakly interacting partons, with the effect of the 
interaction introduced perturbatively in the basis of states. 
Such a starting point for the evaluation of the off-diagonal susceptibility 
is most appropriate for the case of a weakly interacting 
plasma of quarks and anti-quarks, \eg, in the high temperature limit. 
In this weak coupling limit, the expectation of the susceptibility operator 
may be obtained order by order in the strong coupling constant $g$. The 
leading order contribution is  vanishing: 
\bea
\lc \chi_{ij} \rc  = \frac{\lc N_{q_i} N_{q_j} \rc - \lc N_{q_i} \rc \lc N_{q_j} \rc}{VT} \simeq 0 ,
\eea
if  $i \neq j$ as these create and annihilate different flavors. The first non-zero correction to the off-diagonal 
susceptibility at $\mu=0$ occurs at order $g^6$ (see Ref.~\cite{Blaizot:2001vr}).

The opposite case is that of the strong coupling limit, where the system is 
composed of bound states of quarks (or anti-quarks) of flavor $i$ with 
anti-quarks (or quarks) of flavor $j$. The operator of Eq.~\eqref{sus_oprtr} 
may no longer be meaningfully evaluated perturbatively starting from free 
particle states. A convenient starting point is afforded by the basis of 
bound states: in the interest of simplicity we focus on the specific example 
of the low temperature phase of two-flavor QCD, \tie, a system with only 
$u,d$ quarks (anti-quarks) and gluons. At vanishing baryon and charge 
chemical potential, the system may be effectively described in the basis of 
pions $\pi^+,\pi^-,\pi^0$.  As derived in the appendix, in such a system, 
the quark number operators assume the simplified forms, 
\bea
\hat{N}_{u} = \hat{N}_{\pi^+}  -  \hat{N}_{\pi^-}, \nn \\
\hat{N}_{d} = \hat{N}_{\pi^-}  -  \hat{N}_{\pi^+}. 
\label{n_q_to_n_pi}
\eea
In the above, $\hat{N}_{\pi^+}, \hat{N}_{\pi^-}$ is the number operator for 
$\pi^+, \pi^-$. There exists a subtle difference in the meaning of 
$\hat{N}_q$ (where $q$ stands for either quark) and $\hat{N}_{\pi}$: 
the quark number operators are meant to indicate the net amount of a certain 
quark flavor \tie,  $\hat{N}_{u}$ is the difference between the number of 
$u$-quarks and $\bar{u}$ anti-quarks, whereas, the pion number operators 
$\hat{N}_{\pi}$  merely indicate the number of pions of a certain flavor  
and not the difference between the $\pi^+$ and $\pi^-$ populations. 

Within the effective form of  the quark number operators, given by 
Eq.~\eqref{n_q_to_n_pi}, the off-diagonal quark number covariance in 
two-flavor QCD may be constructed as 
\bea 
\sigma^2_{ud} &=& \lc \hat{N}_{u} \hat{N}_{d} \rc - \lc \hat{N}_{u} \rc \lc \hat{N}_{d} \rc \nn \\
&=& \lc  ( \hat{N}_{\pi^+}  -  \hat{N}_{\pi^-} )  ( \hat{N}_{\pi^-}  -  \hat{N}_{\pi^+} )  \rc \nn \\
&-& \lc  \hat{N}_{\pi^+}  -  \hat{N}_{\pi^-} \rc  \lc \hat{N}_{\pi^-}  -  \hat{N}_{\pi^+}  \rc.
\eea
In the case of a dilute pion gas in the  grand canonical ensemble, the 
various flavors may be considered to be uncorrelated. As a result,
\bea
\lc   \hat{N}_{\pi^+} \hat{N}_{\pi^-} \rc \simeq \lc   \hat{N}_{\pi^+} \rc \lc  \hat{N}_{\pi^-} \rc. 
\eea
Incorporating the above approximation in the expression for $\sigma^2_{ud} $, 
leads to the simplified form for the off-diagonal covariance in a dilute 
pion gas at low temperature and vanishing chemical potentials (we also 
denote such contributions as $\sq_{ud}(M)$, where $M$ denotes contributions 
solely from the mesonic sector; at a higher temperature, this will also 
include contributions from more massive mesons \eg, $\ro,\w$ \etc), 
\bea 
\sq_{ud} &=& - \lc {\hat{N}_{\pi^+}}^2 \rc + {\lc \hat{N}_{\pi^+} \rc}^2  
- \lc {\hat{N}_{\pi^-}}^2 \rc + {\lc \hat{N}_{\pi^-} \rc}^2 \nn \\
&=& - (  \sg^2_{\pi^+} + \sg^2_{\pi^+} ) \simeq \sq_{ud}(M).
\eea
As the variance of either pion species is always positive, we obtain the 
general result that at low temperature and vanishing chemical potentials, 
the off-diagonal susceptibility $\chi_{ud} = \sg^2_{ud}/ V$ is negative. 
This prediction has been verified in lattice calculations of $\chi_{ud}$  
in Ref.~\cite{Allton:2005gk}. 

As the temperature of the system is raised, the pion populations 
(and populations of heavier mesons) as well as the fluctuations 
in the populations will increase, leading to a drop in $\chi_{ud}$. 
This trend will continue until substantial baryon populations appear. 
The expression for $\sg^2_{ud}$  in the baryon sector is quite different 
from that in the meson sector, owing to the fact that there are no valence 
anti-quarks in a baryon, nor any valence quarks in an anti-baryon. Hence, 
one obtains $\sg^2_{ud}$ in the baryon sector as 
\bea
\sg^2_{ud}(B) &=&  \lc  ( 3 \hn_{uuu} + 2 \hn_{uud} + \hn_{udd} \nn \\ 
&&-  3\hn_{\ub \ub \ub} - 2 \hn_{\ub \ub \db}  - \hn_{\ub \db \db} )  \nn \\
&\times&  ( 3 \hn_{ddd} + 2 \hn_{udd} + \hn_{uud} \nn \\ 
&&- 3\hn_{\db \db \db} - 2 \hn_{\ub \db \db} - \hn_{\ub \ub \db} ) \rc \nn \\
&-& \lc ( 3 \hn_{uuu} + 2 \hn_{uud} + \hn_{udd} \nn \\ 
&&-  3\hn_{\ub \ub \ub} - 2 \hn_{\ub \ub \db}  - \hn_{\ub \db \db} ) \rc \nn \\
&\times&  \lc ( 3 \hn_{ddd} + 2 \hn_{udd} + \hn_{uud} \nn \\ 
&&- 3\hn_{\db \db \db} - 2 \hn_{\ub \db \db} - \hn_{\ub \ub \db} ) \rc \nn \\
&=& 2 \sq_{uud} + 2 \sq_{udd} + 2 \sq_{\ub \ub \db} + 2 \sq_{\ub \db \db}.
\eea
As a result, the baryon contribution to the off-diagonal susceptibility 
is always positive. As the temperature of the system is raised, the  
contributions from mesons and baryons begin to compensate each other. 
In a weakly interacting hadron gas the two contributions are additive:
$\sq_{ud} =  \sq_{ud} (M) + \sq_{ud}(B)$. The increasing density of 
states in the baryon sector relative to the meson sector at higher 
energies, as well as the larger prefactors involved in $\sq_{ud}(B)$ 
lead to an increasing cancellation between the two contributions as the  
temperature is raised.  

In lattice computations of the temperature dependence of $\chi_{ud}$ one 
notes an initial drop followed by a rise to zero at $T \ra T_c$. If the 
picture of a weakly interacting hadron gas remained valid past $T=T_c$, 
$\sq_{ud}$ would continue to rise to larger positive values. The absence 
of such behavior is an indication of the breakdown of the picture of a 
weakly interacting hadron gas near and beyond $T=T_c$. Further comparisons 
between the behavior of the off-diagonal susceptibility as well as its 
derivatives as computed on the lattice, with expectations within the 
picture of a weakly interacting hadron gas are carried out in the upcoming 
section.

%%%%%%%%%%%%%%%%%%%%%%%%%%%%%%%%%%%%
%%%%%%%%%%%%%%%%%%%%%%%%%%%%%%%%%%%%
%%%%%%%%%%%%%%%%%%%%%%%%%%%%%%%%%%%%
\section{Lattice versus models}
%%%%%%%%%%%%%%%%%%%%%%%%%%%%%%%%%%%%
%%%%%%%%%%%%%%%%%%%%%%%%%%%%%%%%%%%%
%%%%%%%%%%%%%%%%%%%%%%%%%%%%%%%%%%%%

As pointed out in the previous sections, the various susceptibilities and 
their ratios 
may be measured on the lattice \cite{Gavai:2002jt} by  evaluating the 
average of certain operators over a set of configurations. The appropriate 
choice of observables and their sensitivity to composite structures was 
discussed in the previous section. Presently we focus on the results 
obtained from such a calculation on the lattice and use it to isolate 
the subset of models which describe the emergent degrees of freedom 
at various temperatures in strongly interacting matter.  The models used 
will be rather empirical and will require very little beyond arguments 
based on general 
symmetry principles. In all comparisons, the focus will lie expressly on two 
regions: the region below $T_c$ where one expects a hadronic resonance gas 
to be the correct set of degrees of freedom and above $1.5T_c$ where one 
would expect to be firmly in the deconfined phase. 

%%%%%%%%%%%%%%%%%%%%%%%%%%%%%%%%%%%%
%%%%%%%%%%%%%%%%%%%%%%%%%%%%%%%%%%%%
%%%%%%%%%%%%%%%%%%%%%%%%%%%%%%%%%%%%
\subsection{Hadron gas to quasi-particle plasma}
%%%%%%%%%%%%%%%%%%%%%%%%%%%%%%%%%%%%
%%%%%%%%%%%%%%%%%%%%%%%%%%%%%%%%%%%%
%%%%%%%%%%%%%%%%%%%%%%%%%%%%%%%%%%%%

In the region below $T_c$ the monotonic rise of the pressure and energy 
density with temperature, has come to be understood in the picture of a 
weakly interacting hadronic resonance gas \cite{Karsch:2003zq}, as originally
introduced by Hagedorn. This picture remains true for the case of the diagonal 
and off-diagonal susceptibilities.  The susceptibilities are computed, 
assuming the condition of Eqs.~\eqref{sg_qs},\eqref{sg_qs2}, \tie,

\bea 
\chi_{ud} = \frac{\sum_{f} \lc n_f \rc u_f d_f}{VT},
\eea

\begin{figure}
\resizebox{2.7in}{2.7in}{\includegraphics[0in,0in][5in,5in]{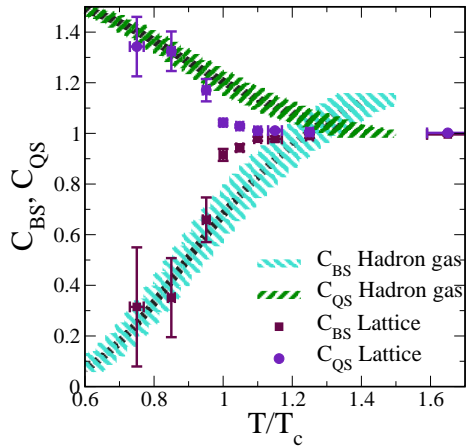}}
\caption{(Color online) A comparison of the $C_{BS}$ and $C_{QS}$ calculated in a truncated hadron 
resonance gas at $\mu_B=\mu_S=\mu_Q=0$MeV compared to lattice calculations at $\mu=0$ from 
Ref.~\cite{Gavai:2005yk}. The two hazed bands for $C_{BS}$ and $C_{QS}$ for the hadron gas 
plots reflect the uncertainty in the actual value of the phase transition temperature $T_c$., which 
is assumed to lie in the range $T_c = 170 \pm 10$MeV. }
\label{fig1b}
\end{figure}

\nt
where, $\lc n_f \rc, u_f, d_f$ are the thermal average, upness and 
downness of a given hadron species $f$. The sum is, in principle, 
over all hadrons, but is usually truncated at an appropriately chosen
upper mass limit.  As a comparison, we plot in Fig.~\ref{fig1b}., 
the coefficients $C_{BS}$ and $C_{QS}$ as obtained from a hadronic 
resonance gas spectrum, truncated at the mass of the $\Omega^-$. As in the case 
of the computation on the lattice, the plots correspond to all chemical potentials 
vanishing. One notes, that the hadron resonance gas provides a good description of 
the behavior of the ratio of susceptibilities  up to the point of the phase transition. 
Here the behavior of the truncated spectrum fails to reproduce the sharp rise in $C_{BS}$  
and the corresponding sharp drop in $C_{QS}$. It should be pointed out that 
while in the lattice simulations, exact isospin symmetry has been imposed, no such 
condition has been required of the hadron spectrum: the masses are taken directly 
form the particle data book. 

The results from the lattice simplify  in the high temperature phase, where a general 
statement regarding susceptibilities may be made:  off-diagonal flavor 
susceptibilities are vanishing compared to the diagonal susceptibilities \cite{Gavai:2002jt,Allton:2005gk};
susceptibilities inclusive of strangeness are smaller compared to those which involve 
lighter flavors. This may be stated as,

\bea
 \chi_{us} = \chi_{ds} \leq \chi_{ud}  \ll \chi_{s} \leq \chi_{d} = \chi_{u},
\eea

\nt
where the last and the first equality applies in the case of isospin symmetry. 
For simulations at vanishing chemical potentials, the mean values of the 
conserved flavor charges are vanishing, as a result $\lc B \rc = \lc S \rc = \lc Q \rc = 0$
Using the above equalities derived form the lattice, we may formulate the 
correlation ratio $C_{BS}$ as
\bea
-3 \frac{\lc B S \rc}{\lc S^2 \rc} 
= \frac{\lc( u + d + s ) s \rc  }{\lc s^2 \rc} \approx  
\frac{\chi_{us} + \chi_{ds} + \chi_s }{\chi_s}  \approx 1.
\label{constraint}
\eea

Models of the deconfined phase must obey the above constraint. The simplest 
model of deconfined matter is that of non-interacting quark, anti-quark and 
gluon quasi-particles. As has been demonstrated in \cite{Koch:2005vg}, in 
such a situation, off-diagonal susceptibilities are identically zero as the 
degrees of freedom may carry only a single flavor. To wit, using the notation 
$  \bar{n}_f  = \lc n_f \rc  + \lc n_{\bar{f}} \rc $, for the total number 
of quarks and anti-quarks of flavor $f$,

\bea
\frac{\chi_{us} + \chi_{ds} + \chi_{s} }{\chi_s} 
&=& \frac{ \bar{n}_u (1\times0) + 
 \bar{n}_d (1\times 0) + \bar{n}_s(1^2)  } {\bar{n}_s(1^2)} \nn \\ 
&=& 1.
\eea

\nt
The zero entries in the above equation indicate that the up and down flavors 
carry no strangeness. We also work in the limit described in Sect.~II,  
where the masses of the quasi-particles are large enough for classical Poisson statistics 
to apply. Thus a model of quark quasiparticles presents a $C_{BS}$ which is in 
agreement with that derived form the lattice. It should be pointed out that the 
nature of the gluon sector is irrelevant in this test. The gluons carry no conserved 
flavor and are thus oblivious to any such constraints. 
As a result, such comparisons yield no clues to the structure of the gluon sector. 

The next independent set of ratios of susceptibilities is that involving the covariances 
of Eq.~\ref{s_qb_and_sb}. Expressions for $C_{SB}$ may be expressed as 
above, 

\bea 
C_{SB} &=&  -3 \frac{\chi_{us} + \chi_{ds} + \chi_{s}}
{ \chi_u + \chi_d + \chi_s + 2\chi_{us} +2 \chi_{ds} + 2\chi_{ud}} \nn \\
&=&  -3 \frac{\chi_{s}}{\chi_u + \chi_d + \chi_s} \geq -1 ,
\eea

\nt
where the last inequality holds in the general case. $C_{SB}=-1$ in the 
case of exact SU$(3)_f$ symmetry, \tie, when mass of the $s$-quark equals the 
mass of its lighter counterparts. Once again, it may be demonstrated 
that the simplified model of quark quasi-particles satisfies this requirement 
for the ratio of susceptibilities. Thus, from this standpoint, it is a viable 
candidate for the degrees of freedom of hot strongly interacting matter. 
Both these conclusions may be reduced to the single observation 
that the degrees of freedom of excited matter have to be such, as to  display 
minimal strength in the covariance between flavors, \tie, off-diagonal 
flavor susceptibilities have to be tiny compared to the diagonal ones. 

%%%%%%%%%%%%%%%%%%%%%%%%%%%%%%%%%%%%
%%%%%%%%%%%%%%%%%%%%%%%%%%%%%%%%%%%%
%%%%%%%%%%%%%%%%%%%%%%%%%%%%%%%%%%%%

\subsection{A plasma with colored meson and diquark bound states?}

%%%%%%%%%%%%%%%%%%%%%%%%%%%%%%%%%%%%
%%%%%%%%%%%%%%%%%%%%%%%%%%%%%%%%%%%%
%%%%%%%%%%%%%%%%%%%%%%%%%%%%%%%%%%%%

The picture of quasi-particle quarks is a rather simple solution to the 
constraint of Eq.~\eqref{constraint}. Indeed, such a picture of a quasiparticle plasma 
has been the subject of numerous studies in weak coupling expansions
\cite{Andersen:2002ey,Blaizot:2003iq,Blaizot:2003tw}.
Such approximations do indeed obtain a similar behavior as a function of 
temperature as compared to the lattice 
for both the off-diagonal and the diagonal susceptibility. 
Recently, an alternate picture of the QGP has been proposed: one where a tower of 
bound states of quarks and gluons are present besides the quasiparticles themselves 
\cite{Shuryak:2004cy,Shuryak:2004tx}. 
The larger number of particles and larger  scattering cross-sections that result 
in such a mixture is shown to account for the pressure observed on the lattice as a 
function of the temperature. The large cross-sections imply very short mean 
free paths and lend consistency to the macroscopic hydrodynamic picture henceforth 
use to  describe the dynamical evolution of the matter produced at RHIC.  

However, in a plasma containing bound states of quarks, in the form of colored 
mesons, diquarks and quark-gluon bound states, the correlation between flavors 
is no longer negligible compared to the diagonal susceptibility. Consider a 
gas consisting solely of quark-antiquark bound states, built from three flavors 
of quarks. The possible states are $u\bar{d},d\bar{u},u\bar{s},s\bar{u},d\bar{s},s\bar{d}$.
The flavor singlets $q\bar{q}$ will be ignored as they carry no conserved flavor and 
thus do not contribute to any susceptibility or covariance. 
In such a system, the ratio $C_{BS}$ must vanish, because all states have vanishing
baryon number. Expressed via the flavor (co-)variances, the numerator of $C_{BS}$ 
is given as 

\bea
\lc (u + d + s) s\rc
&=& \mbox{} - n_{u\bar{s}} - n_{s\bar{u}}  - n_{d\bar{s}} - n_{s\bar{d}}
\nn \\
& +&  n_{u\bar{s}} + n_{s\bar{u}} + n_{d\bar{s}} + n_{s\bar{d}}
\nn \\
&=& 0 .
\eea

\nt
Thus the inclusion of mesonic states changes ratios such as $C_{BS}$ as they cause the 
off-diagonal susceptibilities to become non-vanishing (in this case $\chi_{us} = -\chi_{s}/2$).

Quark gluon bound states contribute similarly as quark-antiquark quasiparticles, 
whereas gluonic bound states make no contribution. States such as diquarks have a 
somewhat opposite effect, diquarks belonging to the flavor anti-triplet ($ud, us, ds$)  
contribute $us = ds = s^2 =+1$, thus leading to (we here assume $\mu_B = 0$):
 
\bea
C_{BS} = \frac{2n_{us} + 2n_{ds} + 2n_{us} + 2n_{ds}}{2 n_{us} + 2 n_{ds}}  = 2,
\eea

\nt
whereas the states belonging to the flavor hexaplet produce a coefficient,

\bea
C_{BS} = \frac{ 2 n_{us} + 2 n_{ds} + 2 n_{us} + 2 n_{ds} + 8 n_{ss} }
              { 2 n_{us} + 2 n_{ds} + 8 n_{ss}   } > 1.0.
\eea

\nt
Ref. \cite{Shuryak:2004cy} provides masses and degeneracies for the various bound states. 
The masses of all the bound states exceed the temperature in this model, allowing us to 
use a Maxwell-Boltzmann (MB) distribution to calculate the populations of such states. 
Such a computation was carried out  in Ref.~\cite{Koch:2005vg} at a temperature of 
$T=1.5T_c$ and yielded a value $C_{BS} = 0.62$ quite different form the value of unity 
found on the lattice.

%%%%%%%%%%%%%%%%%%%%%%%%%%%%%%%%%%%%
%%%%%%%%%%%%%%%%%%%%%%%%%%%%%%%%%%%%
%%%%%%%%%%%%%%%%%%%%%%%%%%%%%%%%%%%%

\subsection{Introduction of baryonic bound states}

%%%%%%%%%%%%%%%%%%%%%%%%%%%%%%%%%%%%
%%%%%%%%%%%%%%%%%%%%%%%%%%%%%%%%%%%%
%%%%%%%%%%%%%%%%%%%%%%%%%%%%%%%%%%%%

These results have motivated the inclusion of a variety of baryonic states into the 
model outlined above \cite{Liao:2005pa}. We illustrate the effect of such additions 
on the correlation between flavors in the following simple model. In the interest of 
simplicity, the flavor group will be restricted to SU(2)$_f$. Lattice results for 
susceptibilities and their derivatives with respect to baryon 
chemical potential using dynamical quarks exist in this case \cite{Allton:2005gk}. 
The model will consist of quark and 
anti-quark quasi-particles $u,d$ and $\bar{u},\bar{d}$, meson like bound states 
$u\bar{u},d\bar{d},u\bar{d},d\bar{u}$, diquark states $uu,dd,ud$ and their antiparticles 
as well as baryons $uuu,uud, udd, ddd$ as well as the corresponding anti-baryons. 
We assume that there is no significant covariance between these quasiparticles, 
which are assumed to be massive enough for MB statistics to apply. We will compute 
general contributions to the off-diagonal susceptibility and its various derivatives. 

In such a situation, the off-diagonal susceptibility at vanishing chemical 
potential may be decomposed as

\bea
\chi_{ud} = \frac{1}{VT} \left[ -2 n^0_{u \bar{d}} + 2 n^0_{u d} + 4 n^0_{uud} + 4 n^0_{udd} \right],
\eea

\nt
where, as before, $2 n^0_x $ includes similar contributions from both particles and 
anti-particles at vanishing chemical potentials. 
It is also assumed that populations of higher excited states, \eg, the hexaplet of diquarks, as 
well as states lying in the baryon decuplet are included in the respective populations.
In the remaining, we will deal with densities as opposed to the absolute numbers: 

\bea
\rho^0_x = \frac{n^0_x}{V}.
\eea

\nt
Given the off-diagonal susceptibility in a range of temperatures, one obtains a 
temperature dependent relation between the baryonic and mesonic densities, \tie,

\bea
2\rho^0_{u \bar{d}}(T) &=& 2\rho^0_{u d}(T) + 4 \rho^0_{uud}(T) + 4 \rho^0_{udd}(T) \nn \\
&-&  
T \chi_{ud} (T, \mu=0), \label{q_bar_q_to _diquarks}
\eea

\nt
where $\rho^0_x(T)$ represent the densities of various quasiparticle species at 
temperature $T$ and vanishing chemical potential.  Unlike the conventional use of the 
term baryonic density, $\rho^0_{uud}$ and $\rho^0_{udd}$ denote the density of a 
certain type of baryon and not the difference between the baryon and anti-baryon 
densities (\tie, $\rho^0_{uud}, \rho^0_{udd}$ do {\em not} denote the net baryon 
density). Adjusting the baryonic densities 
compared to the mesonic densities, one may obtain the requisite 
off-diagonal susceptibility. Introducing a large enough baryon 
density one may engineer a vanishing $\chi_{ud}$ and as an 
extension a vanishing $\chi_{us}$. With such densities, a 
$C_{BS} = C_{QS} = 1$ may also be achieved by a plasma 
of bound states. 

To differentiate a plasma of quasi-particle quarks and anti-quarks 
which naturally produces a $\chi_{ud} \ra 0$ from a plasma of 
colored bound states with a similar property, one needs to 
consider the derivatives of the off-diagonal susceptibility $\chi_{us}$. 
At finite baryon chemical potential, the susceptibility 
will vary, as the populations of  the diquarks and the baryons 
change under the influence of the baryon chemical potential 
(quark anti-quark bound states carry no baryon number and 
hence remain unaffected). One obtains:
\bea
T\chi_{ud}(T,\mu) &\simeq&  -2 \rho^0_{u \bar{d}}(T) 
+ \Large\{ 2 \rho^0_{u d}(T) + 4 \rho^0_{uud}(T) \nn \\
&+& 4 \rho^0_{udd}(T) \Large\} \cosh(\mu \B),  \label{sus_mu}
\eea

\nt
where $\B$ is the inverse temperature. The expression is valid in the 
regime where MB statistics may be used instead of the full quantum 
statistics.  Differentiating Eq.~\eqref{sus_mu}, with respect to 
$\mu \B$, one obtains the relation

\bea
T \left[ \frac{\prt^2 \chi_{ud}}{ \prt (\mu \B)^2} \right]_{\mu = 0} 
&=&  2 \rho^0_{u d}(T) + 4 \rho^0_{uud}(T) + 4 \rho^0_{udd}(T) \nn \\
&=& T \left[ \frac{\prt^4 \chi_{ud}}{ \prt (\mu \B)^4} \right]_{\mu = 0} .
\label{chi2_chi4}
\eea

\nt
The model at this stage is applicable to any 
situation where there are baryonic and mesonic degrees of freedom which 
carry well defined quantum numbers of upness or downness. The baryons and 
mesons may be colored or color singlets. The considerations outlined 
above are applicable to all such models, including the hadron resonance 
gas model used below $T_c$. 

\begin{figure}
\resizebox{2.7in}{2.7in}{\includegraphics[0.5in,0.5in][5in,5in]{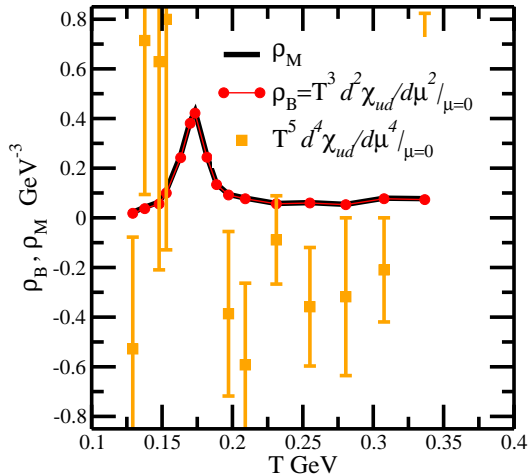}}
\caption{(Color online) A plot of the mesonic and baryonic contributions to $\chi_{ud}(T,\mu=0)$ and 
and a plot of the derivatives $T\prt^2 \chi_{ud} / \prt (\mu \B)^2|_{\mu=0}$ (which 
are equal to the baryonic contribution $\rho_B$) and $T\prt^4 \chi_{ud} / \prt (\mu \B)^4|_{\mu=0}$ 
(which is negative beyond $T_c = 170$MeV and thus is inconsistent with a bound 
state interpretation of the data.). See text for details.}
\label{fig3}
\end{figure}

In this way, one may divide the contributions to the off-diagonal 
susceptibility and its derivatives in terms of mesonic and baryonic 
contributions. Using the measured susceptibility and its derivatives, 
these contributions may be estimated as a function of the temperature.
In the lattice computations, results are expressed in units of $T_c$; 
we assume $T_c = 0.17$ GeV for definiteness.
These are plotted as the thick solid line (mesons) and the solid circles 
(baryons)  in Fig.~\ref{fig3}. These densities for 
$\rho_B = 2 \rho^0_{u d}(T) + 4 \rho^0_{uud}(T) + 4 \rho^0_{udd}(T)$ and
$\rho_M = 2 \rho^0_{u \bar{d}}(T)$ satisfy Eq.~\eqref{q_bar_q_to _diquarks} 
and the first equality of Eq.~\eqref{chi2_chi4}.
The condition imposed by Eq.~\eqref{chi2_chi4} 
has to be satisfied by the derivatives of the susceptibility in such a 
picture of bound states. The fourth derivative of the susceptibility 
has been plotted as the square symbols. Despite large error bars one 
notes that, while the baryon density or the second derivative of the 
susceptibility is consistent with Eq.~\eqref{chi2_chi4} below the phase 
transition temperature, it becomes inconsistent with such a condition 
above the phase transition temperature. Above $T_c$, 
$T \prt^4 \chi_{ud} / \prt (\mu \B)^4$ is actually negative; hence, 
no composite quasiparticle picture is compatible with such results. 
It has already been pointed out in Ref.~\cite{Allton:2005gk} that the 
signs of the various derivatives of the susceptibility are consistent 
with the picture of a weakly interacting quasi-particle gas. 

In the above simple model, numerous approximations were made. Maxwell-Boltzmann statistics 
was used throughout, variances were replaced with the mean populations of the various flavors, 
and masses of various flavors were assumed to be independent of chemical potential. 
Such approximations were made to clearly illustrate the central point of this article, 
that the measured values of the off-diagonal susceptibilities and their derivatives are 
inconsistent with a picture of a composite quasiparticle plasma. On the other hand,
the susceptibility in a weakly interacting  model of quark quasi-particles, computed 
in the hard-thermal loop approximation has been shown to be consistent with the 
diagonal, and off-diagonal susceptibilities derived from lattice simulations
\cite{Blaizot:2001vr}. It has also been pointed out that the signs of the derivatives of 
the susceptibilities are consistent with the quasi-particle picture \cite{Allton:2005gk}. 
A computation of the absolute values of  the derivatives in such a picture is currently 
underway.

\section{Experimental observables}

In the previous sections, a theoretical study of various diagonal and off-diagonal 
susceptibilities has been carried out and their relations with the degrees of freedom 
in heated strongly interacting matter has been elucidated. In the present section, 
our focus will lie on the possible measurement of such correlations in heavy-ion 
experiments. Our considerations will be restricted to the measurement of the ratio 
$C_{BS}$ which in the view of the authors is the most favorable from an experimental
point of view. 

It is believed that thermalized, strongly interacting and deconfined matter is 
transiently produced in central heavy ion collisions at RHIC. If one divides the 
whole system into small rapidity bins, then the 
fluctuations of conserved charges within a given rapidity bin are controlled by the 
degrees of freedom prevalent at the temperatures achieved. As the system expands and 
cools, it reconverts to a hadronic gas prior to freeze-out. If  the transition to the 
confined phase is sudden, as in the case of a continuous transition and the longitudinal 
expansion is sufficiently large, then the net charge in the rapidity bin, set in the 
deconfined phase is maintained through the hadronic phase up to freezeout. 
Such fluctuations may then be measured event by event. The two major hurdles in the survival 
of such fluctuations through the hadronic phase are the contamination by hadronic fluctuations and 
measurability of the particles sensitive to the partonic fluctuations.  The first issue 
is off lesser importance for an observable such as $C_{BS}$ as the lightest 
carriers of strangeness are the kaons which are much heavier than the temperatures 
reached in RHIC collisions in the hadronic phase. They are produced in far fewer 
number than pions and hence do not manage to diffuse through multiple rapidity bins in the 
short time available in the hadronic phase. The measurement of baryon number is the 
primary problem in the estimation of $C_{BS}$. 

The detector most suitable to measurements of bulk fluctuations at RHIC is the STAR detector. 
In the measurement of baryon-strangeness correlations, the detector has to accurately assess the 
baryon number and strangeness in a given rapidity bin in each event. As the STAR detector 
is blind to stable uncharged particles it cannot measure the neutron and antineutron populations. 
As a result, a measurement of $\sg_{BS}$ may become rather difficult. 
Based on the discussion of Sect.~II, we present the following recourse. A new quantum 
number is constructed,

\bea
M = B + 2 I_3,
\eea

\nt
and fluctuations of $M$ with respect to $S$ are studied. In theoretical calculations of the 
the quantity $\sg_{MS}$ as outlined in Sects. II and IV, one notes that the assumption of
isospin symmetry (see Eq.~\eqref{isospin}) 
reduces the covariance $\sg_{MS}$ to simply $\sg_{BS}$, \tie,

\bea
\sg_{MS} &=& \lc (B + 2 I_3) S \rc   - \lc B+2I_3\rc \lc S\rc \nn \\
&=& \sg_{BS} + 2 \sg_{I_3 S} \simeq \sg_{BS}. \label{ms=bs}
\eea

\nt
As a result, in all theoretical models with isospin symmetry $C_{MS} = C_{BS}$. 

In the experimental determination, $M$ has the advantage that
it is vanishing for all particles that do not carry charge or strangeness,  thus 
$M=0$ for neutrons, antineutrons, neutral pions, \etc. The experimental measure is 
thus 

\begin{equation}
C_{MS}\ =  - 3 \ { \sum_n M^{(n)} S^{(n)} - 
\left(\sum_n M^{(n)}\right) \left(\sum_n S^{(n)}\right) 
\over \sum_n (S^{(n)})^2 - \left(\sum_n S^{(n)}\right)^2}\ .
\end{equation}

\nt
In the above equation $M^{(n)}$ and $S^{(n)}$ are the total 
$M$ and total strangeness within the given rapidity bin in event $(n)$. 
One may not make the simplification of counting the product quantum number $MS$ for 
individual flavors as in Eq.~\eqref{sg_qs} as the fluctuations in $M$ and $S$ are 
set in the partonic phase and the final hadrons are the result of decay from the 
deconfined phase. Hence,  the different flavors are no longer uncorrelated as 
assumed in the derivation of Eq.~\eqref{sg_qs}.

The presence of $I_3$ in the observable, introduces a new problem in the 
experimental measurement. The lightest 
carriers of $M$ are the charged pions which are numerous in the hadronic 
phase and may lead to contamination of the conserved charge in the chosen rapidity 
bin from neighboring bins.  However, as the central rapidity bins at RHIC are 
practically charge neutral, the possibility of contamination by charged pion
fluctuations is greatly reduced. One may divide  the measured correlation 
between $I_3$ and $S$  into a genuine correlation and a contamination,  

\bea 
\sg_{I_3 S} = \sg_{I_3 S}^{\rm act} + \sg_{I_3 S}^{\rm cont}. \label{pi_cont}
\eea

\begin{figure}
\resizebox{2.5in}{2.5in}{\includegraphics[0.5in,0.5in][5in,5in]{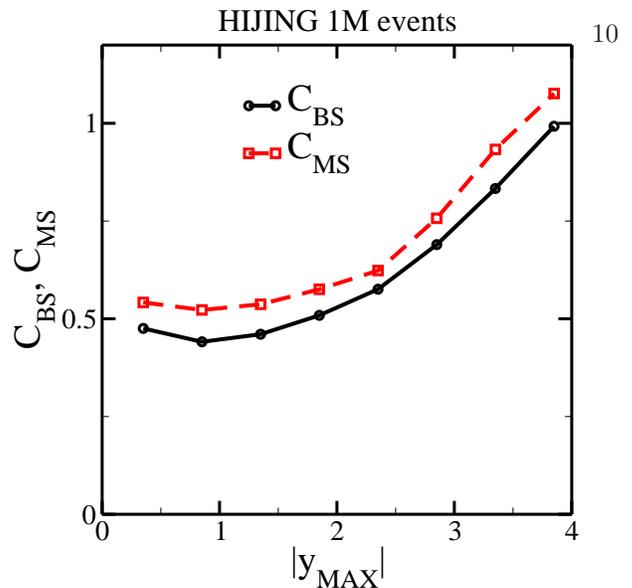}}
\caption{(Color online) A comparison of the two related ratios of variances $C_{BS},C_{MS}$ as a 
function of  the acceptance in rapidity from $-|y_{max}|$ to $|y_{max}|$.}
\label{fig4}
\end{figure}

\nt
As the fluctuations that result in $\sg_{I_3 S}^{\rm cont}$  are driven 
by pions in the hadronic phase, a sum over a relatively large number of 
events will lead to this quantity becoming rather small compared to the 
actual correlation $\sg_{I_3 S}^{\rm act}$ if violations of isospin symmetry are 
negligible. This condition should hold for the produced hadronic phase over a 
range of rapidities at RHIC.  This effect is illustrated 
in Fig.~\ref{fig4} where both $C_{BS}$ and $C_{MS}$ are calculated from  
model simulations using the HIJING code \cite{Wang:1991ht}.

In Fig.~\ref{fig4} the correlations $C_{BS}$ and $C_{MS}$ are estimated in a central 
$Au-Au$ event at $\sqrt{s} = 200$ $A$GeV. 
The acceptance in rapidity ranges from $-|y_{max}|$ to 
$|y_{max}|$, hence a larger $y_{max}$ indicates a larger acceptance. 
In an effort to further mimic the experimental acceptance, $K_L$ mesons are 
ignored, and $K_S$ mesons are identified either as a $K^0$ or a $\bar{K^0}$ 
with 50\% probability for either case. 
The results are presented as a function of $y_{max}$. One notes 
that 
% up to a 
over the range of $y_{max}$ the two correlations $C_{BS}$ and $C_{MS}$ are
rather similar. 
% At larger rapidities, the increasing baryon number 
% introduces corrections to the isospin symmetry assumed in Eq.~\eqref{ms=bs}. 
% The baryonic matter at larger rapidities carries net isospin and this leads to an 
The increased fluctuations in isospin are the cause of the slightly larger value of  $C_{MS}$  as 
compared to $C_{BS}$.
% for $y_{max} > 1.0$. 
This bodes well for the measurement of $C_{BS}$  in 
RHIC experiments via a measurement of the quantity $C_{MS}$, over a range of 
rapidity intervals.  
% with the possible
% limitation that the measurement may be restricted to the rapidity range between 
% $-1$ to 1.

\section{Conclusions}

Both heavy ion collisions and lattice simulations of QCD at finite temperature 
present components in the study of heated strongly interacting matter. In this 
article we have demonstrated that off-diagonal susceptibilities and ratios of 
susceptibilities have the ability to discern the 
prevalent flavor carrying degrees of freedom in heated strongly interacting matter. 
The latter quantity  may be measured on the lattice as well as in heavy-ion collisions, 
under the assumption that event-by-event fluctuations in a heavy-ion collision are 
set in the deconfined phase and are maintained through the hadronic phase. 

In this paper, the behavior of a number of observables based on the 
ratio of susceptibilities $C_{BS}, C_{QS}$, \etc\ was explored
both in the confined as well as the deconfined phase. Under the assumption 
of isospin symmetry, simplifying relations between such observables were 
derived, which reveal the interdependence of such ratios, 
\eg, Eqs.~(\ref{cqs_cbs},\ref{cqb_csb}). Results of the computations 
of such quantities in a hadronic resonance gas as well as in a non-interacting 
plasma of quarks and gluons were compared with calculations on the lattice (Fig.~\ref{fig1b}).
Such comparisons demonstrate that the flavor carrying sector of QCD, is consistent with a 
deconfined plasma of weakly interacting quarks and anti-quarks above $T_c$. Below $T_c$ 
the behavior of $C_{BS}$ and $C_{QS}$ is consistent with that of a hadron resonance gas. 

The various relations between the ratios of susceptibilities relate the behavior of $C_{QS}$ 
to that of $C_{BS}$. The behavior of $C_{BS}$ above and below $T_c$ is caused primarily by the 
vanishing of the the off-diagonal susceptibility $\chi_{us}=\chi_{ds}$ at $T\geq T_c$ 
and the negative value of its expectation below $T_c$. A similar behavior is shown by other 
off-diagonal susceptibilities such as $\chi_{ud}$. The remainder of our study focused on 
the behavior of the two flavor off-diagonal susceptibility $\chi_{ud}$, as calculations of 
the temperature dependence of $\chi_{ud}$ as well as its various derivatives have been 
carried out in full unquenched lattice simulations. 

In Sects. III and IV, it was demonstrated that the behavior of $\chi_{ud}$ as well as its 
various derivatives is consistent with that of a hadron gas below $T_c$ and a weakly 
interacting plasma of quarks and anti-quarks above $T_c$. The behavior of the various 
derivatives of $\chi_{ud}$ with baryon chemical potential was shown to be inconsistent 
with a bound state picture above $T=T_c$. Such an inconsistency remained even if 
baryonic bound state populations above $T_c$ were artificially enhanced to be consistent 
with the $C_{BS}$ and $C_{QS}$ measured on the lattice. 

Finally, we have proposed new experimental observable $C_{MS}$ related by isospin symmetry 
to $C_{BS}$. This observable is blind to uncharged and non-strange particles. We showed 
that it is equivalent to $C_{BS}$ and may be measurable in experiments at RHIC. Estimates 
of $C_{MS}$ and $C_{BS}$ in HIJING simulations demonstrate the similarity of the two 
quantities over a range of rapidities at RHIC. This bodes well for its use 
as an experimental proxy for $C_{BS}$. Such measurements offer the possibility to 
directly probe the degrees of freedom in the deconfined matter produced in high-energy 
heavy-ion collisions. 

{\em Acknowledgments:}  This work was supported in part by the U.S. Department of Energy
under grant DE-FG02-05ER41367. The authors thank S. Bass for helpful discussions.  
A. M. thanks V. Koch and J. Randrup for stimulating collaboration, part of the results of 
Sect. III represent an alternate derivation of the results to appear in Ref.~\cite{KMR06}.

\section{appendix}

In this appendix, we outline the derivation of Eq.~\eqref{n_q_to_n_pi}. 
Imagine strongly interacting matter at low temperature, confined in a 
box of volume $V$. The temperature is assumed to low enough for the prevalent 
degrees of freedom to be a dilute gas of pions. The state vector representing a pion with momentum $p = \frac{2 n_p \pi }{ V^{1/3}}$ 
may be expressed as 

%\bea
%
%|\pi^+_p \rc &=& \int d q \!\!\!\!\!\!\! \sum_{  \mbox{\tiny $ \left( \begin{array}{*{3}{c@{,}}c@{;}}
%n_1 & n_2 & \cdots & n_{\infty} \\
%\bar{n}_1 & \bar{n}_2 & \cdots & \bar{n}_{\infty} \\
%m_1 & m_2 & \cdots & m_{\infty} \\ 
%\bar{m}_1 & \bar{m}_2 & \cdots & \bar{m}_{\infty}
%\end{array} \right) $  } }
%
 %\!\!\!\!\!\!\!\Psi^q {\mbox{ \tiny $ \left( \begin{array}{*{3}{c@{,}}c@{;}*{3}{c@,}c@{;}}
%n_1 & n_2 & \cdots & n_{\infty} & \bar{n}_1 & \bar{n}_2 & \cdots & \bar{n}_{\infty} \\
%m_1 & m_2 & \cdots & m_{\infty} & \bar{m}_1 & \bar{m}_2 & \cdots & \bar{m}_{\infty}
%\end{array} \right) $  }}  \nn \\ 
%
%&\times&  \mbox{ $ \left| \begin{array}{*{3}{c@{,}}c@{;}*{3}{c@,}c@{;}}
%n_1 & n_2 & \cdots & n_{\infty} & \bar{n}_1 & \bar{n}_2 & \cdots & \bar{n}_{\infty} \\
%m_1 & m_2 & \cdots & m_{\infty} & \bar{m}_1 & \bar{m}_2 & \cdots & \bar{m}_{\infty}
%\end{array} \right\rangle $  } \nn \\
%
%&\times& \kd \left( \sum_{i}  n_i - \sum_{i}  \bar{n}_i - 1 \right) 
%
%\kd \left( \sum_{i}  \bar{m}_i - \sum_{i} m_i - 1 \right)  \nn \\
%&\otimes& \Phi^{p-q} | g_{p-q} \rangle
%\eea

\bea
|\pi^+_{\vec{p}} \rc &=& \sum_{   \{ {\bf n} \}, \{{ \bf \bar{n}} \} , \{ \bf {m} \} , \{ \bf {\bar{m}}  \} , \{ {\bf l} \}   }  
\!\!\!\!\!\!\!\!\!\!\Psi^{\vec{p} }_1 (      \{ {\bf n} \}, \{{ \bf \bar{n}} \} , \{ \bf {m} \} , \{ \bf {\bar{m}}  \} , \{ {\bf l} \}      ) \nn \\
&\times&  |      \{ {\bf n} \}, \{{ \bf \bar{n}} \} , \{ \bf {m} \} , \{ \bf {\bar{m}}  \}         \rc \nn \\
&\times& \kd \left( \sum_{i}  n_i - \bar{n}_i - 1 \right) 
\kd \left( \sum_{i}  \bar{m}_i -  m_i - 1 \right)  \nn \\
&\otimes&  |   \{ {\bf  l} \}  \rc . \label{one_pi_state}
\eea

\nt
In the above equation, the vectors of integers 
$\{ {\bf n} \} (\{{ \bf \bar{n}} \}), \{ {\bf  m} \} (\{ {\bf \bar{m}}  \})$, represent the 
set of occupation numbers in different momentum states of $u$-quarks ( $\bar{u}$-anti-quarks) and 
$d$-quarks ( $\bar{d}$-anti-quarks), \tie,

\bea
\{ {\bf n} \} \equiv \{ n_1, n_2, n_3, \cdots  \}.
\eea

\nt
Values for $n_i$ may be 0 or 1. The vector $\{ {\bf l} \}$ represents the occupation 
numbers of the gluon sector and is a vector of  integers $l_i\geq 0$. 
In Eq.~\eqref{one_pi_state}, 
$ |      \{ {\bf n} \}, \{{ \bf \bar{n}} \} , \{ \bf {m} \} , \{ \bf {\bar{m}}  \}         \rc $ represents 
the general state vector of the quark (anti-quark) sector.  
The function $ \Psi^{\vec{p}}_1 (      \{ {\bf n} \}, \{{ \bf \bar{n}} \} , \{ \bf {m} \} , \{ \bf {\bar{m}}  \},   \{ {\bf l} \}   )$ 
represents the wave function of the one pion state with the constraint that the total momentum residing 
in this sector is $\vec{p}$. Hence,

\bea 
 && \Psi^{\vec{p}}_1 (  \{ {\bf n} \}, \{{ \bf \bar{n}} \} , \{ {\bf m} \} , \{ { \bf \bar{m} } \}, \{ {\bf l} \} )   \\
&=&
\widetilde{\Psi}  (  \{ {\bf n} \}, \{{ \bf \bar{n}} \} , \{ \bf {m} \} , \{ \bf {\bar{m}}  \} ,  \{ {\bf l} \}  )   \nn \\
&\times& 
 \kd \left( \vec{p} - \sum_{i}  \vec{p}_{u,i} n_i +  \vec{p}_{\bar{u},i} \bar{n}_i  +  \vec{p}_{d,i} m_i  
+ \vec{p}_{\bar{d},i} \bar{m}_i  + \vec{p}_{g,i} l_i  \right)  , \nn
\eea

\nt
where, $\vec{p}_{u,i}$ is the momentum of the $i$th $u$-quark state with occupation $n_i$. 
While not explicitly pointed out,  the wave function $\Psi^{p}_1$ also maintains over all color neutrality.

Given the form of the one pion state, it is a trivial matter to formulate general expressions for 
multiple pion states. In this way, an effective basis of states at low temperature is constructed:
$
| 0 \rc , | \pi^+_{p_1} \rc , | \pi^+_{p_1}  \pi^+_{p_2}  \rc, | \pi^+_{p_1}  \pi^-_{p_2}  \rc
$
etc. Interactions between these various states, is assumed to be small enough to be estimated in a 
perturbative formalism. The reader will note that the various states outlined above are 
orthogonal, given the orthogonality of the various states in the quark gluon occupation number basis 
used in Eq.~\eqref{one_pi_state}.
Such $n$-pion states may also be expressed in an occupation number basis as above, e.g.,

\bea
| \pi^+_{p_1} \rc \equiv | 0_1, 0_2, \cdots, 0_{p_1 - 1},1_{p_1},  0_{p_1 + 1}, \cdots \rc .
\eea

\nt
One may now express the quark number operators as a matrix in the occupation number basis of pion states, \tie, 
\bea
&& \hat{N}_u = \sum_{ \{ { \bf n } \} , \{ { \bf m} \} }   \label{qrk_num_in_pi_basis}\\
&& |n_1,n_2, \cdots \rc \lc n_1,n_2, \cdots |  \hat{N}_u |m_1,m_2,\cdots \rc  \lc m_1,m_2, \cdots |  \nn
\eea 
\nt
Using the expression for the one pion state from Eq.~\eqref{one_pi_state}, we obtain the simple 
relation, 

\bea
\hat{N}_u  | 0_1, 0_2, \cdots, 0_{p_1 - 1},1_{p_1},  0_{p_1 + 1}, \cdots \rc . &=& 
\hat{N}_u |  \pi^+_{p_1} \rc  \nn \\
&=&  |  \pi^+_{p_1} \rc 
\eea 
\nt
Similarly, the action of the upness operator on the one $\pi^-$ state may be computed to 
be 

\bea
\hat{N}_u |  \pi^-_{p_1} \rc = - |  \pi^-_{p_1} \rc .
\eea

Generalizing to the $n$-pion states, one obtains the general relations for the quark 
number operators in the basis of pions (in the 
limit that the pion gas is dilute \tie, the interactions between the different states are small ),

\bea
\hat{N}_{u} = \hat{N}_{\pi^+}  -  \hat{N}_{\pi^-}, \nn \\
\hat{N}_{d} = \hat{N}_{\pi^-}  -  \hat{N}_{\pi^+} . \label{n_q_to_n_pi_A}
\eea

\end{document}